\documentclass[prb,twocolumn,superscriptaddress,amsmath]{revtex4}
\usepackage{graphicx}
\usepackage{epsfig}
\usepackage{float}
\usepackage{amsmath}
\usepackage{amssymb}
\begin{document}
\title{Implementation and testing of Lanczos-based algorithms for
  Random-Phase Approximation eigenproblems.}

\author{Myrta Gr\"uning}
\affiliation{Centre for Computational Physics and Physics Department,
  University of Coimbra, Rua Larga 3004-516 Coimbra, Portugal}
\affiliation{European Theoretical Spectroscopy Facility, NAPS/IMCN, Universit\'e Catholique de Louvain, B-1348
Louvain-la-Neuve, Belgium}
\author{Andrea Marini}
\affiliation{European Theoretical Spectroscopy Facility, Physics Department,
   University `Tor Vergata", I-00133 Rome, Italy}
\affiliation{Nano-Bio Spectroscopy Group, Universidad del Pa\'\i s Vasco,  E-20018 San Sebasti\'{a}n, Spain}
\affiliation{IKERBASQUE, Basque Foundation for Science, E-48011 Bilbao, Spain}
\author{Xavier Gonze}
\affiliation{European Theoretical Spectroscopy Facility, NAPS/IMCN, Universit\'e Catholique de Louvain, B-1348
Louvain-la-Neuve, Belgium}

\begin{abstract}
The treatment of the Random-Phase Approximation Hamiltonians, encountered in different frameworks, like Time-Dependent Density Functional Theory or Bethe-Salpeter equation, is complicated by their non-Hermicity. Compared to their Hermitian Hamiltonian counterparts, computational
methods for the treatment of non-Hermitian Hamiltonians are often less
efficient and less stable, sometimes leading to the breakdown of the
method. Recently [Gr\"uning et al. Nano Lett. {\bf 8}, 2820  (2009)], 
we have identified that such Hamiltonians are usually pseudo-Hermitian.
Exploiting this property, we have implemented an algorithm of the
Lanczos type for random-Phase Approximation Hamiltonians that benefits
from the same stability and 
computational load as its Hermitian counterpart, and applied it to
the study of the optical response of carbon nanotubes. We present here
the related theoretical grounds and technical details, and study the
performance of the algorithm for the calculation of the optical absorption of a
molecule within the Bethe-Salpeter equation framework. 
\end{abstract}
\maketitle

\section{Introduction}
The Random-Phase Approximation (RPA) Hamiltonian $H^{\text{RPA}}$
appears in several areas of physics and theoretical chemistry, and describes
strong collective excitations of a many-body system as the linear
combination of particle-hole pairs $|\lambda \mu\rangle$~\cite{FetterW71,RingS80}. It has the
form 
\begin{equation}
  \label{eq:rparpa}
   H^{\text{RPA}} = 
\begin{pmatrix}
 R  & C \\
-C^{*}  & -R^*
\end{pmatrix},
\end{equation}
where the resonant $R$ and anti-resonant $-R^*$ blocks are Hermitian
matrices in the subspace generated by
particle-hole pairs propagating respectively forward ($|\lambda \mu\rangle$)
and backward ($|\tilde{\mu \lambda \rangle}$) in time (in what follows
$\alpha,\lambda$ indicate particles and $\beta,\mu$ holes),
and the $C$ and $-C^*$ blocks are symmetric matrices coupling  the
particle-hole pairs propagating forward and backward in time.
The excitation energies and strengths of the many-body system are the
eigensolutions of Eq.~(\ref{eq:rparpa}). 
Note that the RPA Hamiltonian is not Hermitian, thus its
eigenvalues are not necessarely real. 
  
In quantum chemistry, condensed matter physics, nanoscience, or nuclear physics, the RPA Hamiltonian appears within the
state-of-the-art approaches for calculating the excitations in an
electronic system: the time-dependent density functional theory~\cite{RungeG84}
(TD-DFT) and the Bethe-Salpeter~\cite{SalpeterB51} (BS)
equation.~\cite{endnote27} 
In the commonly used approximations to TD-DFT (e.g. real exchange-correlation kernel) and 
BS equation (static screening of the interaction), 
{\it all the eigenvalues are real}.
TD-DFT is particularly successful for finite systems, namely
molecules and molecular clusters, while the BS approach is mostly
used for extended systems, like periodic bulk solids and, in general, for
systems where excitonic effects play an important role~\cite{OnidaRR02}. 
Nowadays, the application of these approaches to the computation of the
time-dependent responses of more and more complex
systems, such as large bio-molecules or nanostructures,  poses the
problem of efficient solution of 
the eigenproblem for $H^{\text{RPA}}$. 
For large matrices, the direct
diagonalization is usually not possible, and one has to resort to
iterative algorithms, such as the Lanczos method. Such algorithms
exist for both Hermitian and non-Hermitian Hamiltonian. However,  
compared to their Hermitian Hamiltonian counterparts, algorithms
for the treatment of non-Hermitian Hamiltonians are often less
efficient and less stable, sometimes leading to the breakdown of the
method~\cite{Bai00,Tretiak09}.

Within TDDFT, the very convenient Hermitian formulation of the eigenvalue
problem proposed by Casida~\cite{Casida95} exists. However its
application is limited to 
finite systems and purely local effective potentials for which the
$H^{\text{RPA}}$ is real. The presence of e.g. spin-orbit coupling prevents
the application of  Casida's approach.
In general a further
approximation is introduced, the so-called Tamm-Dancoff approximation
(TDA), that considers only particle-hole pairs propagating forward in
time, so that the TDA Hamiltonian corresponds just to the resonant part,
$H^{\text{TDA}} = R$.
The TDA is often sufficiently
accurate, as in the case of optical absorption spectra of
periodic bulk systems. On the other hand, the TDA becomes inaccurate or even
unphysical in the case of electron-energy-loss
spectra~\cite{OlevanoR01}, reflectivity spectra~\cite{MarinopolousG10}, and also for the
optical absorption of low-dimensional systems, e.g. nanosystems or
$\pi-$conjugated molecules~\cite{GruningMG09,MaRM09}.
In a previous work~\cite{GruningMG09} we have implemented an approach for the
solution of the RPA Hamiltonian, that avoids the TDA, and still
benefits of the efficiency and robustness of the algorithms for the
Hermitian case. 
This approach has been already successfully applied  to
the calculation of the optical absorption and energy-loss spectra of a carbon nanotube.
While our previous work focussed on the implications of the TDA for
nanoscale systems, in this work the focus is on the theoretical grounds and some more technical
aspects of that approach. We show here how the Lanczos algorithm for Hermitian
eigenproblem (Sec.~\ref{sc:lncmet}) can be used for the RPA Hamiltonian, that is
pseudo-Hermitian with real eigenvalues (Sec.~\ref{sc:psdhmc}), by simply redefining
the inner product (Sec.~\ref{sc:innprd}). We obtain
(Sec.~\ref{sc:newlnc}) the generalization to complex matrices of
the scheme proposed by Van der Vorst in Ref.~\onlinecite{VdVorst82}.
The obtained algorithm is then further specialized (Sec.~\ref{sc:lncepm}) to the
calculation of the macroscopic dielectric function (from which the
optical absorption and energy-loss spectra are derived)  and finally
applied to the calculation of the optical response of the
trichloro-bezene isomers within the BS equation framework (Sec.~\ref{sc:resmol}), to show the algorithm
accuracy (Sec.~\ref{sc:resacc}) and efficiency (Sec.~\ref{sc:reseff}).

\section{Mathematical background}~\label{sc:teobck} 
This section reviews briefly the two key ``ingredients'' of the presented
approach: the Lanczos method for the solution of (non-)Hermitian
eigenproblems, and the definition of pseudo-Hermitian matrix. The
Lanczos method allows to calculate by recursion the eigenvalues, and
eigenvectors, or directly the response spectrum, of large matrices. The
pseudo-Hermicity is related to the reality of the eigenvalues of a
matrix and with the possibility of transforming the matrix into a
Hermitian matrix.
\subsection{Lanczos method for Hermitian eigenproblems}~\label{sc:lncmet} 
The Lanczos recursion method~\cite{Bai00} is a general algorithm 
for solving eigenproblems for a Hermitian operator $H$.
This algorithm recursively builds an orthonormal basis
$ \{| q_i \rangle \}$ (Lanczos basis) in which $H$ is represented as a
real symmetric tridiagonal matrix,
\begin{equation}\label{eq:tridia}
T^k=\left(
\begin{matrix}
a_1    &  b_2   & 0       &  \cdots & 0 \\
b_2    &  a_2   & b_3     &         & \vdots \\
0      & \ddots & \ddots & \ddots  &  0\\
\vdots &        & b_{k-1} & a_{k-1}  & b_{k} \\
0      & \cdots & 0      & b_{k}   & a_k \\
\end{matrix}
\right).
\end{equation}
The first vector $| q_1 \rangle$ of the Lanczos basis is set equal to
a (normalized) given vector $| u_0\rangle / \|u_0\|$. The next vectors
are calculated from the three-term relation
\begin{equation}\label{eq:lnczos}
|Q_{j+1}\rangle = H |q_j\rangle - a_j | q_j\rangle -
b_j | q_{j-1}\rangle,   
\end{equation}
where
\begin{align}\label{eq:abxprs}
a_j = \langle q_j| H |q_j\rangle, \\
b_{j+1} = \|Q_{j+1}\|, \\
|q_{j+1} \rangle =|Q_{j+1} \rangle / b_{j+1}.  
\end{align}
%%%
\begin{figure}[h]
\begin{center}
\includegraphics[clip,width=8cm]{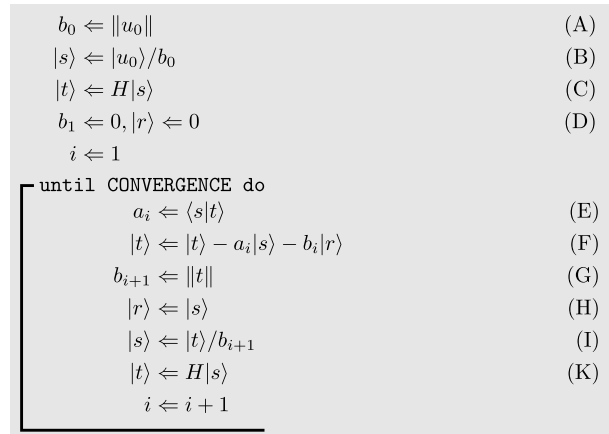}
\end{center}
\vspace{-.5cm}
\caption{Hermitian Lanczos algorithm}
\label{fg:lnczsH}
\end{figure}
%%%
The algorithm is schematically described in Fig.(\ref{fg:lnczsH}). In
steps (A)-(D) the variables are initialized before entering the
conditional loop [steps (E)-(K)]. Here, at each iteration a new vector
of the Lanczos basis is computed till the convergence criteria is met.
The cost per iteration is given mainly
by the matrix-vector multiplication at step (K), that is of $O(N^2)$ for non-sparse matrices, with
$N$ the size of $H$. In terms of memory and storage, if one is just interested in the eigenvalues, at each
iteration only three vectors ($|q_{n-1} \rangle, |q_{n} \rangle,
|q_{n+1} \rangle$) are needed, and only two reals ($a_i,b_i$) need to be
stored. At the end of the
process one gets the tridiagonal matrix of
Eq.~(\ref{eq:tridia}) of dimension $k \times k$, that can be
diagonalized with a cost $\propto k$.   
Compared with the standard diagonalization, the advantages are the memory usage, and the computational cost  $\propto kN^2$ (for diagonalization is
$O(N^3)$) as soon as the number of iterations $k \ll N$. This is in
practice always the case when we are interested only in a portion of
the spectrum of $H$.\cite{Grosso95} 

As first highlighted by Haydock~\cite{Haydock80,Cini07}, an additional advantage of Lanczos recursive approach is the possibility of calculating the resolvent $(\omega -H)^{-1}$ matrix elements, bypassing completely the diagonalization. 
In fact the resolvent for the state $|u_0\rangle$ takes the form of a continued fraction
\begin{equation}\label{eq:cntfrc}
\langle u_0|(\omega -H)^{-1} |u_0\rangle = \|u_0\|^2 \cfrac{1}{(\omega - a_1) - \cfrac{b_2^2}{(\omega - a_2) -
    \cfrac{b_3^2}{\dots}}}.   
\end{equation} 
Other matrix elements can be then calculated by recursion (see ~\ref{ap:offdia}). 

\subsection{Lanczos method for non-Hermitian eigenproblems}~\label{sc:lncmnH} 
The Lanczos recursive approach can be extended to the non-Hermitian
case~\cite{Bai00}. For a non-Hermitian matrix $H$, that we suppose
diagonalizable, the action on a ket $|v\rangle$ differs
from the action on a bra $\langle v|$:  
no orthogonal basis set exists, that could transform
it into a diagonal form.
The most straightforward extension of the Lanczos procedure illustrated
in the previous subsection is the Arnoldi recursive approach that
transforms $H$ into an upper-Hessenberg matrix, instead of  a
tridiagonal one, and thus presents clear computational disadvantages
with respect to the Hermitian case.~\cite{endnote28}

It is still possible to tridiagonalize $H$, by defining a
bi-orthonormal 
Lanczos basis $\{\langle p_i|, |q_i\rangle\}$, that is $\langle
p_i|q_j\rangle = \delta_{ij}$ while in general, $\{\langle p_i|\}$,
$\{|q_i\rangle\}$ are not orthogonal. In this basis, $H$ is represented
as a non-Hermitian tridiagonal matrix 
\begin{equation}\label{eq:tdianH}
T'^j=\left(
\begin{matrix}
a_1    &  b_2   & 0       &  \cdots & 0 \\
c_2    &  a_2   & b_3     &         & \vdots \\
0      & \ddots & \ddots & \ddots  &  0\\
\vdots &        & c_{j-1} & a_{j-1}  & b_{j} \\
0      & \cdots & 0      & c_{j}   & a_j \\
\end{matrix}
\right).
\end{equation}
The first vectors, $\langle p_1|$, $| q_1 \rangle$ of the Lanczos basis are set equal to
 given bi-orthonormal vectors $\langle w_0| $, $| u_0\rangle $. The next vectors
are calculated from the three-term relations
\begin{align}\label{eq:lnczsn}
|Q_{j+1}\rangle =& (H - a_j )|q_j\rangle - c_j | q_{j-1}\rangle.\\   
\langle P_{j+1}|=& \langle p_j|(H - a^*_j) - \langle p_{j-1}| b_j .
\end{align}
where
\begin{align}\label{eq:abcxpr}
a_j = \langle p_j| H |q_j\rangle, \\
b_{j+1} = \|Q_{j+1}\|, \\
c_{j+1} = \langle P_{j+1} | Q_{j+1}\rangle/b_{j+1}, \\
\langle p_{j+1}|  =\langle P_{j+1}| / c_{j+1}, \\ 
|q_{j+1} \rangle =|Q_{j+1} \rangle / b_{j+1}.\label{eq:abcxprE}  
\end{align}
%%%
\begin{figure}[h]
\begin{center}
\epsfig{figure=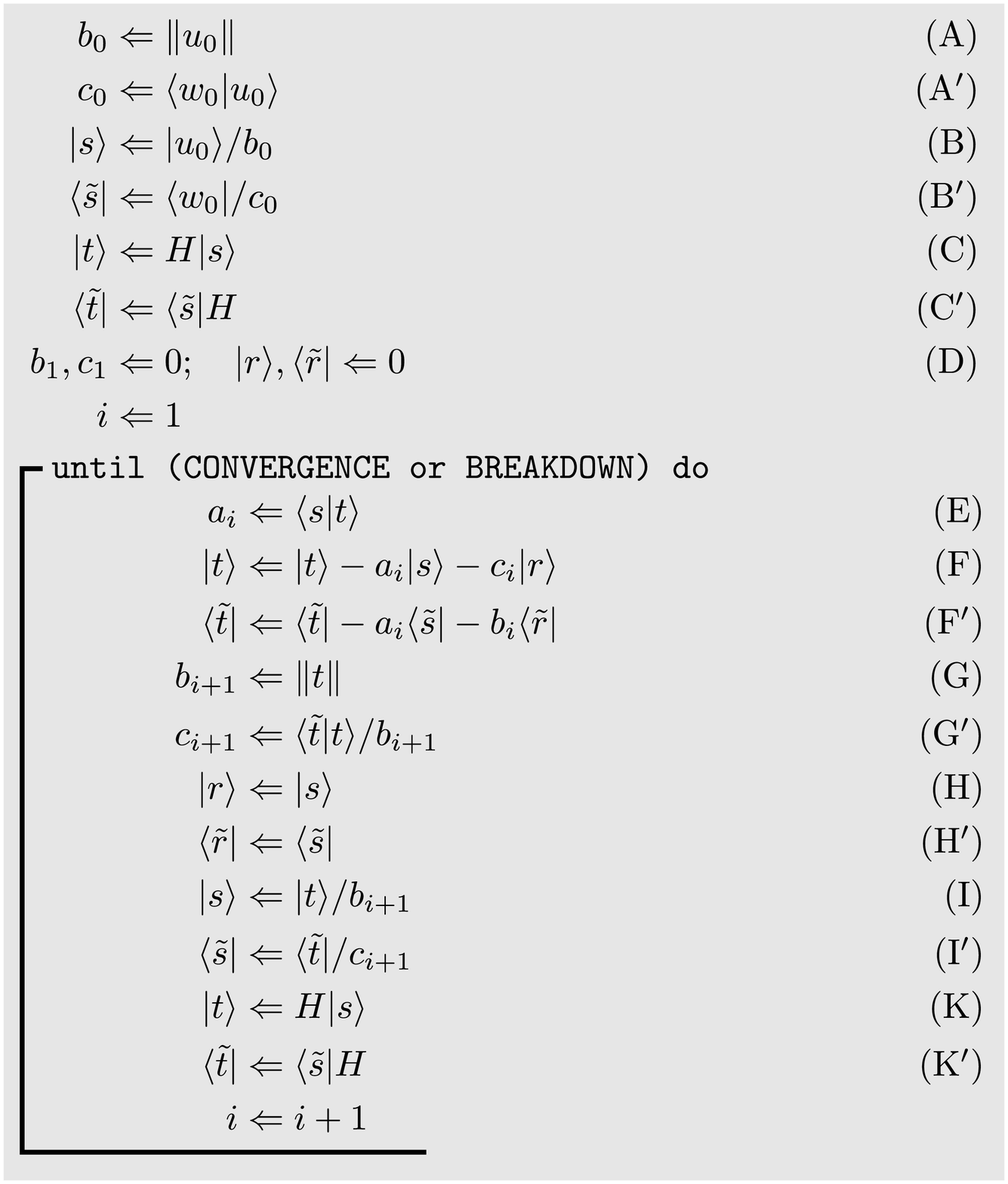,width=8cm}
\end{center}
\vspace{-.5cm}
\caption{Non-Hermitian Lanczos algorithm [Eqs.~(\ref{eq:lnczsn}-\ref{eq:abcxprE})]. The steps that are added
  with respect to the Hermitian case in Fig.~(\ref{fg:lncznH}), are tagged
  with a primed letter. See text for further explanations.}
\label{fg:lncznH}
\end{figure}
%%%
Figure~\ref{fg:lncznH} schematically describes the algorithm
corresponding to Eqs.~(\ref{eq:lnczsn},\ref{eq:abcxpr}). The number of 
steps is doubled with respect to the Hermitian case, since we 
build two sets of basis vectors, so that each operation performed for $|
q_i \rangle$, has to be repeated for the $\langle p_i|$. Another
important difference with respect to the Hermitian case is that, since    
 $\{\langle p_i|\}$, $\{|q_i\rangle\}$ are not orthogonal, it can happen
that one of the new vectors that are built in Eqs.~(\ref{eq:lnczsn}) are
zero or their vector product is zero. 
Then at each iteration, one has to check, together with the
convergence conditions, that the algorithm does not
break down.\cite{endnote29}

Also the Lanczos-Haydock (LH) procedures, Eq.~(\ref{eq:cntfrc}), can be generalized ($b_i^2$ are
replaced by $c_ib_i$). Indeed, the non-Hermitian LH has been applied to the calculation of the
dynamical polarizabilities of molecules within the time-dependent
density functional and Bethe-Salpeter perturbation theory schemes introduced in Refs.~\onlinecite{Walker06,Rocca08,Rocca10}. 

\subsection{Pseudo-Hermicity}~\label{sc:psdhmc}
The Hermicity of an operator $H$, that is $H=H^\dagger$, insures the
reality of the spectrum of $H$. However, Hermicity is a sufficient,
but not necessary condition for the reality of the spectrum of an
operator. 
To study systematically non-Hermitian Hamiltonian with real spectrum,
quite recently, Mostafazadeh~\cite{Mostafazadeh02a} introduced the concept of pseudo-Hermicity. The
$\eta$-pseudo-Hermitian adjoint of $H$ is defined as
\begin{equation}\label{eq:defpsH}
H^{\sharp} :=
\eta^{-1} H^{\dagger} \eta  
\end{equation}
where $\eta$ is an invertible transformation in Hilbert space.
Then, a Hamiltonian is $\eta$-pseudo-Hermitian if $H^{\sharp}
=H$. It is easy to recognize that this definition includes
Hermicity ($\eta = I$). 

Using this concept, it is possible to
define a sufficient as well as necessary condition for the reality
of the spectrum of a matrix. 
It can be proved~\cite{Mostafazadeh02b} that $H$
diagonalizable has a real spectrum if and only if
$H$ is $\eta$-pseudo-Hermitian and there exists an operator $O$ with $\eta = O O^{\dagger}$.  
Since $\eta$ is positive-definite one can define an inner product,
\begin{equation}
\langle \cdot | \cdot \rangle_\eta :=
\langle \cdot | \eta  \cdot \rangle, 
\end{equation}
and the corresponding norm $\| \cdot \|_\eta$.
Then, with respect to this inner product,  $H$ is
Hermitian (alternatively  $O$ transforms $H$ into an Hermitian
Hamiltonian). 

\section{Pseudo-Hermicity of the RPA Hamiltonian}\label{sc:teores}
In this section we first show that the RPA
Hamiltonian is pseudo-Hermitian, and introduce an operative way to
define the inner product with respect to which $H^{\text{RPA}}$ is
Hermitian. As a consequence, the Lanczos algorithm for Hermitian operators presented in the previous
section is simply reformulated using this inner product. Finally the algorithm is specialized
for TD-DFT and BS calculations of excited state properties of electronic systems.   

\subsection{Redefinition of the inner product for RPA Hamiltonian}\label{sc:innprd}  
The RPA Hamiltonian $H$ can be written as~\cite{Zimmermann70,RingS80}
\begin{equation}
  \label{eq:decrpa}
   H = F\bar H =
\begin{pmatrix}
1  & 0 \\
0  & -1
\end{pmatrix}
\begin{pmatrix}
R  & C \\
C^{*}  & R^*
\end{pmatrix}.
\end{equation}
In Eq.~(\ref{eq:decrpa}), $F$ is a Hermitian involution, and $\bar H$
a Hermitian matrix (indeed $R$ is Hermitian, and $C$ is symmetric). $H$ is
pseudo-Hermitian with respect to both $F$ and $\bar H$. In fact, since
$F$ is an involution, $\bar H = F H$, and the $F$-pseudo-Hermicity,
$FH = H^\dagger F$ follows directly from the Hermicity of $\bar
H$. Note that the $F$-pseudo-Hermicity corresponds to the invariance
of the Hamiltonian with respect to $FK$, with $K$ is the complex
conjugation matrix, or more physically to the invariance of the Hamiltonian
under the combined action of parity ($F$) and time-reversal ($K$)
operator. However, since $F$ is clearly non positive definite, it
cannot be used for defining a norm.

More interestingly, the $\bar H$-pseudo-Hermicity is also easily proved
\begin{align*}
  \bar H H = &F H H &&\text{($\bar H$ definition)} \\
  =& H^\dagger F H &&\text{($\bar H$ $F$-pseudo-Hermicity) } \\
  =& H^\dagger \bar H &&\text{($\bar H$ definition)} \quad \square. 
\end{align*}
The positive definitiveness of $\bar H$ is intimately related to the
nature of the independent particle solution $|\Phi\rangle$ from which
the particle-hole pairs are built. In fact $\bar H$ is the curvature
tensor, or stability matrix, that one obtains when expanding the
energy surface around the stationary point $|\Phi\rangle$ for small
variations $|\delta \Phi\rangle$. If $|\Phi\rangle$ corresponds to
a minimum in the energy surface, then $\bar H$ is positive
definite.~\cite{RingS80} For excitations of an electronic system, in connection with
the singlet-triplet instability, Zimmermann~\cite{Zimmermann70} found that a sufficient
condition for $\bar H$ to be positive definite is the smallest
particle-hole pair energy 
being larger than the modulus of the largest matrix element of the
coupling matrix $C$. 

Except in very rare cases, in practical applications the $|\Phi\rangle$ corresponds to a minimum in the energy surface,
hence we can use $\bar H$ to define a modified inner product, with respect to which $H$ is Hermitian
 \begin{equation}\label{eq:sclprd}
\langle \cdot | \cdot \rangle_{\bar H} :=
\langle \cdot | \bar H  \cdot \rangle, 
\end{equation}
and the associate norm. With this inner product, the closure relation
for a complete set of functions $\{\Psi_n\}$, orthonormal with
respect to the modified inner product, has to be defined
accordingly as
\begin{equation}
  \label{eq:cmprel}
  \sum_n \bar H | \Psi_n \rangle \langle \Psi_n | = I.
\end{equation}

\subsection{Lanczos recursive method for RPA Hamiltonians}  \label{sc:newlnc}
Using the properties in previous section, we can redesign the Lanczos
recursive procedure (Sec.~\ref{sc:lncmet}) simply by replacing the
standard inner product with the one defined in
Eq.~(\ref{eq:sclprd}). The result is the algorithm proposed in
Ref.\onlinecite{VdVorst82} generalized to complex matrices.  
The main question is whether the replacement of the inner product is
convenient numerically, with respect to the standard non-Hermitian Lanczos method. Indeed the new inner product implies a matrix-vector multiplication by $\bar H$, and thus, as in the non-Hermitian Lanczos, the computational cost is doubled. However, taking into account that $\bar H$ and $H$ are related by $F$,  effectively only one  matrix-vector multiplication is needed as in the Lanczos method for Hermitian matrices. In fact,
\begin{align}
  a_j = \langle q_j| H q_j\rangle_{\bar H} =&  \langle q_j| H^\dagger F H
  |q_j\rangle =  \langle (H q_j)| F |(H q_j)\rangle,
  \label{eq:acoeff} \\
  b_{j+1} = \|Q_{j+1}\|_{\bar H} =& \sqrt{\langle Q_{j+1}| F  |(H Q_{j+1})\rangle},~\label{eq:bcoeff} 
\end{align}
so that only one matrix-vector multiplication by $H$ appears, while computing the
matrix elements of $F$ will be unexpensive, because
$F$ corresponds to the identity in the subspace of particle-hole
propagating forward in time and to
minus the identity in the subspace of particle-hole
propagating backward in time. 
\begin{figure}[h]
\begin{center}
\epsfig{figure=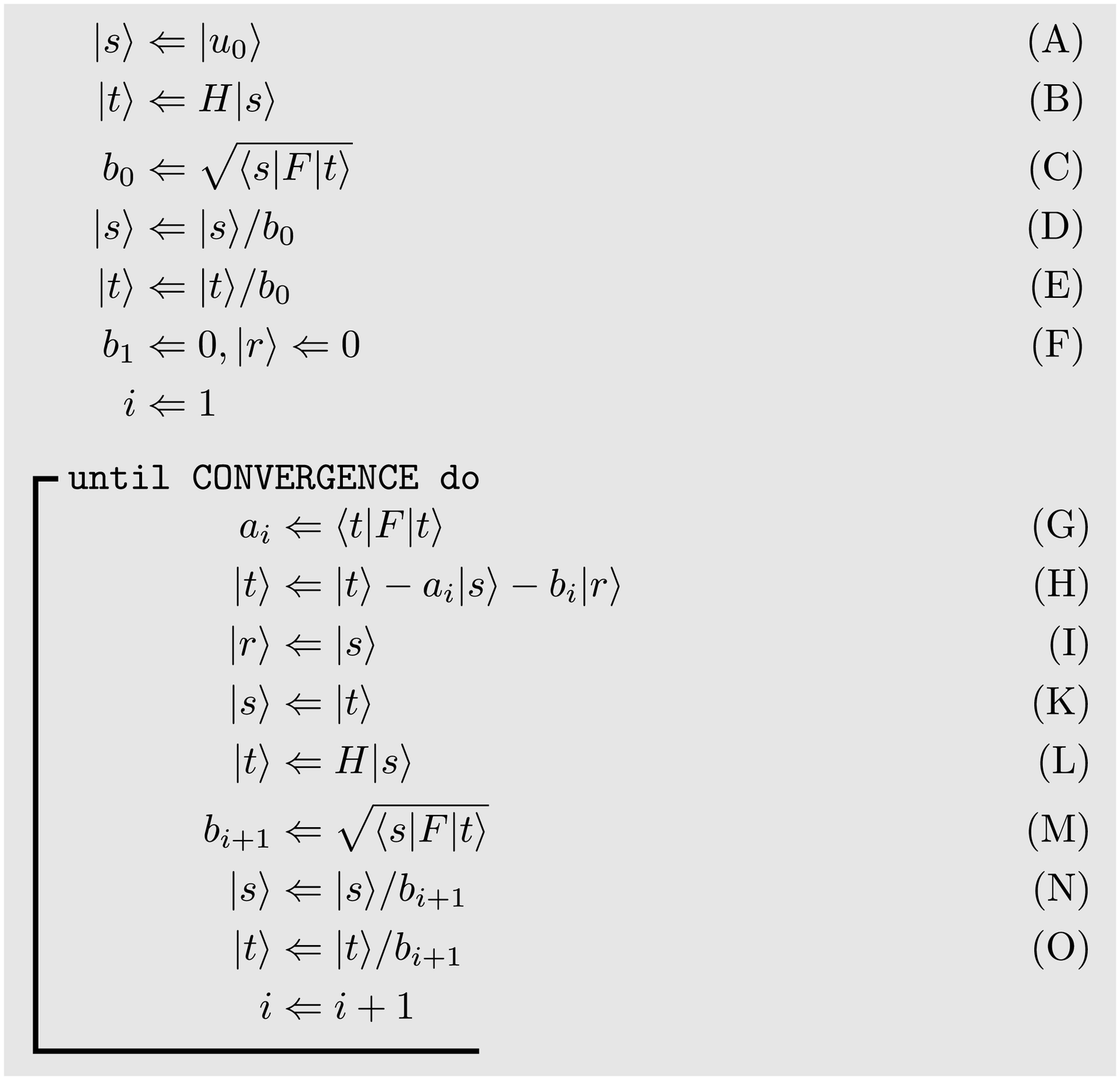,width=8cm}
\vspace{-.5cm}
\end{center}
\caption{Pseudo-Hermitian Lanczos algorithm}
\label{fg:lncPTH}
\end{figure}
The Lanczos algorithm so modified is shown in
Fig.~\ref{fg:lncPTH}. With respect to the Hermitian case both in the
variable initialization [steps (A) to (F)], and in the conditional
loop [steps (G) to (O)] the algorithm is rearranged: the $b_i$ factors
[steps (C) and (M)] are calculated after the matrix-vector
multiplication [steps (B) and (L)]; as a consequence two additional
steps are added to normalize the Lanczos basis vectors [steps (D)-(E),
  (N)-(O)]. The reordering, and addition of the extra (computationally
inexpensive) steps avoids the additional matrix-vector multiplication
due to the modified inner product. The cost is then, as in the Hermitian case, of one vector-matrix multiplication.
Coming to the diagonal matrix elements of the resolvent for the state $|u_0 \rangle$ [Eq.(~\ref{eq:cntfrc}) in the Hermitian case], the straightforward generalization (that is substituting the standard with the $\bar H$-inner product) gives
\begin{equation}\label{eq:cntfrP}
\langle u_0|(\omega -H)^{-1} u_0\rangle_{\bar H} = \|u_0\|_{\bar H}^2 \cfrac{1}{(\omega - a_1) - \cfrac{b_2^2}{(\omega - a_2) -
    \cfrac{b_3^2}{\dots}}}.   
\end{equation}   
However, we are still interested in the matrix element
calculated with the standard inner product. Using the closure relation
[Eq.~(\ref{eq:cmprel})],  we expand $\langle w_0|(\omega -H)^{-1}
|u_0\rangle$ in terms of of  $\langle q_i|(\omega -H)^{-1} |u_0\rangle_{\bar H}$,
\begin{equation}\label{eq:cntfr2}
\langle w_0|(\omega -H)^{-1} u_0\rangle=  \sum_i \langle w_0| q_i\rangle \langle q_i|(\omega -H)^{-1} u_0\rangle_{\bar H}. 
\end{equation}
The first element of the sum $\propto \langle u_0|(\omega -H)^{-1}
u_0\rangle_{\bar H}$ is given by Eq.~(\ref{eq:cntfrP}), while the
other elements are found by recursion (see ~\ref{ap:offdia}). 
The standard product $\langle w_0| q_i\rangle$, can be computed at
each iteration of the Lanczos process and stored. The elements with
$i\ne 1$ are different from zero, since the Lanczos basis vectors are
orthonormal with respect to the $\bar H$-inner product, not with
respect the standard one. On the other hand for the cases we studied
we found that in fact the summation in  Eq.~(\ref{eq:cntfr2}) converges rapidly. 

\subsection{Special case of calculation of the macroscopic dielectric function}\label{sc:lncepm}
The Lanczos approach introduced in the previous subsection can be applied to calculate the macroscopic dielectric function
$\epsilon_M(\omega)$ within either the BS or TD-DFT framework. As shown 
in Ref.~\onlinecite{BenedictS99} the $\epsilon_M(\omega)$ can be rewritten as 
\begin{equation}\label{eq:epsm02}
\epsilon_M(\omega) = 1 - \langle P | (\omega - H)^{-1} F | P \rangle,
\end{equation}
where $| P \rangle$ is a ket whose components in the $|\lambda \mu
\rangle$ and $|\mu\lambda\rangle$ space are the optical oscillators: $\langle P | \lambda\mu\rangle \propto \langle \lambda | \vec{d}\cdot\vec\xi|\mu \rangle$, with 
$\vec{d}$ the electronic dipole, and $\vec\xi$ the light polarization 
factor. 
The components of the $|P\rangle$ vector in a particle-hole
pair basis have the following symmetry 
\begin{equation}\label{eq:Pkprop1}
\langle \mu\lambda | P \rangle= \left( \langle \lambda\mu | P \rangle\right)^*.
\end{equation}
Eq.~(\ref{eq:epsm02}) is straightforwardly calculated from
Eq.~(\ref{eq:cntfr2}) choosing $|u_0 \rangle = F|P\rangle$.
The components of $|u_0 \rangle$ 
vector in a particle-hole
pair basis have the following symmetry 
\begin{equation}\label{eq:Pkprop}
\langle \mu\lambda | u_0 \rangle= -\left( \langle \lambda\mu | u_0 \rangle\right)^*.
\end{equation}

The symmetry property of $|u_0\rangle$ [Eq.~(\ref{eq:Pkprop})] can be used to further reduce the
computational load of the pseudo-Hermitian Lanczos algorithm. Within the
vector space spanned by the particle-hole pairs  let's consider the
subspace ${\mathcal V_-}$ of  
vectors with the property Eq.~(\ref{eq:Pkprop}), and the subspace of
vectors ${\mathcal V_+}$ with the property Eq.~(\ref{eq:Pkprop1}). 
The two subspace are clearly non-overlapping. It can be verified
that $\bar H$ transforms a vector $v\in V_{+(-)}$ in a vector
$w$ belonging to the same subspace, while $F$ transforms a vector
$v\in V_{+(-)}$ in a vector $v'$ belonging to the other subspace. As a
consequence $H$ transforms a vector $v\in V_{+(-)}$ in a vector $w'$
belonging to the other subspace. Also, we note that when $q_j$ belongs
to either one of the subspace, the product in Eq.~(\ref{eq:acoeff}) is
zero. 

Next, using Eq.~(\ref{eq:lnczos}) it can be demonstrated by induction that if
the first Lanczos basis vector $|q_1\rangle$ belongs to one of the subspaces,
the vectors of the Lanczos basis set belongs to either one of the
subspaces depending on the parity of the iteration $j$. More
specifically if---as it is our case---$|q_1\rangle \in {\mathcal V}_-$, then all
$|q_j\rangle$ with $j$ odd belong to ${\mathcal V}_-$, and all $q_j$ with $j$
even belong to ${\mathcal V}_+$.
 In fact, if $q_1\in {\mathcal  V}_-$ then $q_2\in {\mathcal
   V}_+$ (base case). Furthermore, suppose that in Eq.~(\ref{eq:lnczos}), $|q_j\rangle \in {\mathcal
  V}_-$, and $|q_{j-1}\rangle \in {\mathcal V}_+$, then $Q_{j+1}$ (thus the
normalized $q_{j+1}$) belong to ${\mathcal V}_+$ and $Q_{j+2}$
[obtained by applying again Eq.~(\ref{eq:lnczos}) after updating $|q_{n-1}\rangle$,$|q_{n}\rangle$,$|q_{n+1}\rangle$] to ${\mathcal V}_-$ (inductive step).

\begin{figure}[h]
\begin{center}
\epsfig{figure=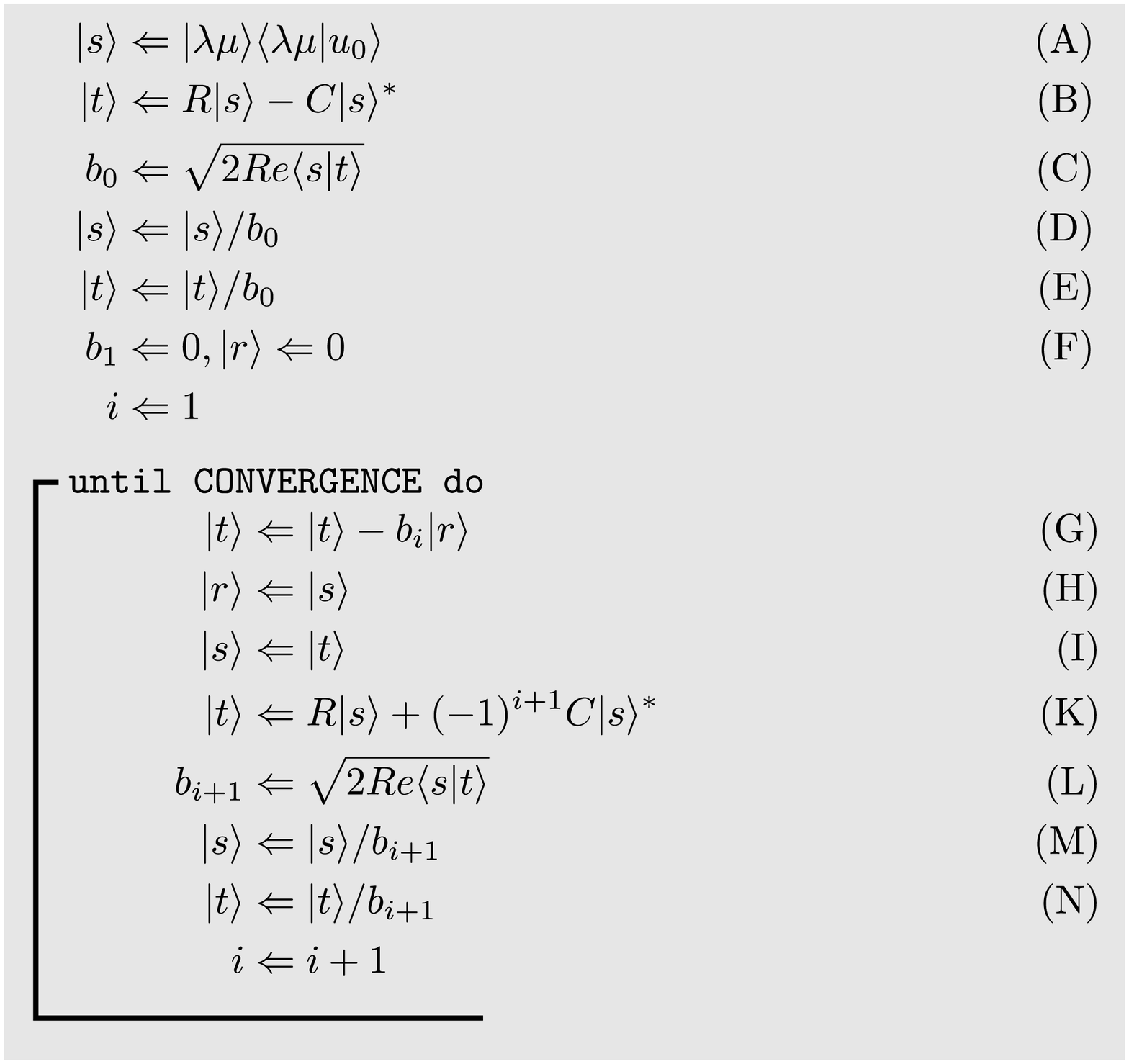,width=8cm}
\vspace{-.5cm}
\end{center}
\caption{Pseudo-Hermitian Lanczos algorithm modified to account for a starting
vector $|P\rangle$ with the anti-Hermitian property given by Eq.~(\ref{eq:Pkprop})}
\label{fg:lncEMH}
\end{figure}

Figure~\ref{fg:lncEMH} shows the modified Lanczos algorithm for the case in
which the starting vector has the symmetry property given by
Eq.~(\ref{eq:Pkprop}).  Since half of the components of $|u_0\rangle$ contains all the
information, the first Lanczos vector [step (A)] is just the projection
$|u_0\rangle$ on the $|\lambda\mu\rangle$ subspace (that is only
particle-hole pairs propagating forward in time). Then, if $H^{\text{RPA}}$ is a $N\times N$
matrix, the algorithm works with vectors and matrices of dimension $N/2$ 
and $N/2\times N/2$ respectively, like in the case of the TDA approximation.
The multiplication by $H^{\text{RPA}}$, that is the operation determing the cost of
the algorithm, is replaced by two matrix-vector multiplications for
matrices of half the size [steps (B) and (K)],
so that the cost is reduced by a factor two compared to the full
matrix case.~\cite{endnote30} 

Similarly, the $\bar H$-norm of a vector
defined in Eq.~(\ref{eq:bcoeff}) is rewritten as in steps (C) and (L)
of Fig.~\ref{fg:lncEMH}.
Furthermore, since all $a_i$ are zero by symmetry they do
not need to be calculated and can be omitted [step (G)] in the
three-terms relation [Eq.~(\ref{eq:lnczos})].

\section{Results}\label{sc:resmol}
We applied the algorithms described in Figs.~(\ref{fg:lnczsH})
and~(\ref{fg:lncEMH}) to the calculation of the optical
absorption of the isomers of the trichlorobenzene (TCB) molecule within the BSE framework.
First, we analyze the effect of the TDA on the spectra of the
isomers, then we assess the accuracy and efficiency of the
algorithms.
%%%%
\subsection{Effect of the particle-hole hole-particle coupling}\label{sc:effcpl}
%%%%
Within the BSE framework the $R$ and $C$ matrix elements are defined as 
\begin{align}\label{eq:bsemte}
R_{\alpha\beta,\lambda\mu} &= (E_{\beta} -
E_{\alpha})\delta_{\alpha\lambda}\delta_{\beta\mu} +
K_{\alpha\beta,\lambda\mu}, \\
C_{\alpha\beta,\mu\lambda} &= K_{\alpha\beta,\mu\lambda}, 
\end{align}
where $E_\alpha$,$E_\beta$ are the (quasi)particle/hole energies. The BS kernel (see e.g. Ref.~\onlinecite{OnidaRR02}) 
\begin{equation}\label{eq:bsekrn}
K_{\alpha\beta,\lambda\mu} = \bar v_{\alpha\beta,\lambda\mu} -W_{\alpha\beta,\lambda\mu},
\end{equation}
describes the interaction between the particle-hole pairs in terms of
particle-hole exchange $\bar v_{\alpha\beta,\lambda\mu}$ ($\bar v$ is the
Hartree potential without the long-range component) and attraction $W_{\alpha\beta,\lambda\mu}$ ($W$ is
the screened interaction). The particle-hole exchange is responsible
of the local-field effects, that is the effect of the induced
microscopic electric fields. The particle-hole attraction
introduces instead the so-called excitonic effects. 

In our calculations the basis of particle-hole pairs $\{|\gamma\lambda\rangle\}$ has been obtained
from the solution of the Kohn-Sham equation for the system. The
Kohn-Sham orbital energies has been corrected to obtain the      
(quasi)particle/hole energies $E_{\alpha}$, $E_{\beta}$ by using the $GW$ approximation. 
The Kohn-Sham calculations have been performed using the
pseudo-potential plane-wave code {\sc abinit}~\cite{abinitref}, while the $GW$ and BSE
calculations have been performed using the {\sc yambo} code~\cite{yamboref} where the
algorithms described in this work have been implemented.
\begin{figure}[H]
\begin{center}
\includegraphics[clip,width=7cm]{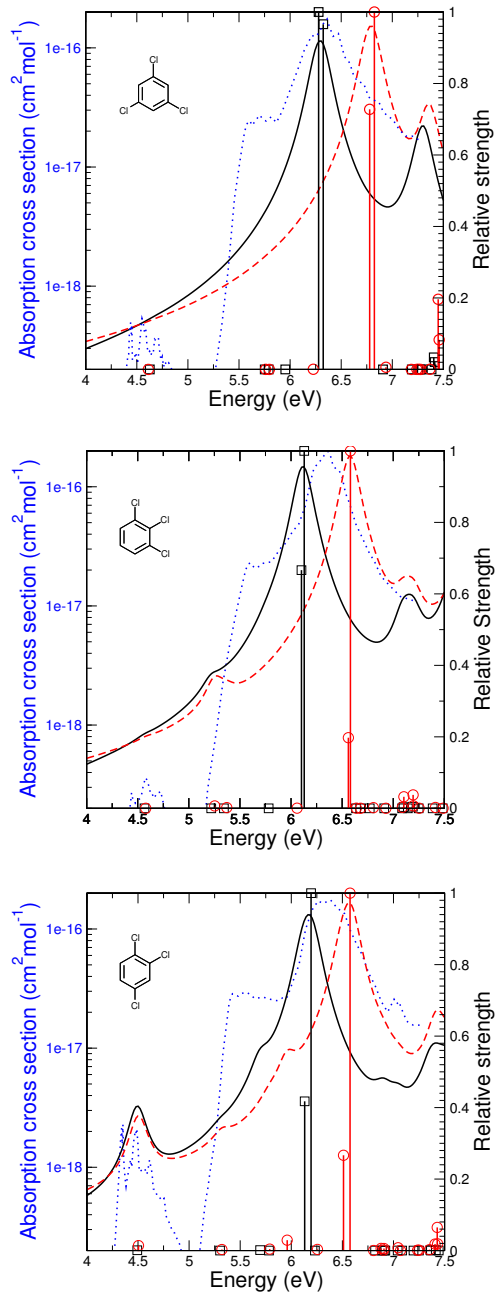}
\end{center}
\vspace{-.5cm}
\caption{[Color online] Optical spectra of TCB isomers: 1,3,5-TCB (top
  panel), 1,2,3-TCB (middle panel),  1,2,4-TCB (bottom panel). The
  imaginary part of the macroscopic dielectric function (in arbitrary
  unit) calculated within BSE using both full (black solid line) and
  TDA (red dashed line) are compared with the gas-phase experimental
  absorption cross section spectra (blue dotted line). The energy and
  relative strength of the excitations for full BSE (black squares)
  and TDA-BSE (red circles) obtained by exact diagonalization are
  reported in linear scale (the largest intensity is normalized to 1).}
\label{fg:CompSp}
\end{figure}
Figure \ref{fg:CompSp} shows the optical absorption spectra of
1,3,5-TCB, 1,2,3-TCB and 1,2,4-TCB within the BSE obtained both by LH
and exact
diagonalization of either the TDA or the full Hamiltonian. 
Results are compared with 
the gas-phase experimental absorption cross section.~\cite{Sharping87} 
The experimental spectra for the three isomers are very similar to
each other with the main peak centered at about 6.3 eV, a shoulder at about 5.5 eV and a very weak feature at 4.5 eV. For 1,2,4-TCB, the isomer with less symmetry, the peak at 4.5 eV is enhanced, the shoulder is broader and more pronounced and an extra peak is visible at 7 eV. 
The full BSE spectrum reproduces fairly well the position of the peaks
in the three isomers. 1,3,5-TCB, 1,2,3-TCB show a dark transition at
4.5 eV, that acquires strength for the 1,2,4-TCB. In correspondence of
the 5.5 eV shoulder in the experimental spectra, 
all the theoretical spectra present excitations. In the 1,3,5-TCB these are dark,
while for 1,2,3-TCB and 1,2,4-TCB a shoulder appears in the spectra. The theoretical spectrum of 1,2,4-TC shows also the additional peak around 7 eV.
The neglection of the coupling within the TDA 
blue-shifts by about 0.5 eV the main peak for all the isomers, with
the effect of worsening the agreement with the experimental data,
especially for what concern the relative position of the peaks. 
The main peak is due to two very close excitations due
mostly to the $\pi\rightarrow \pi^*$ transitions between the doubly degenerate highest
occupied (HOMO) and the doubly degenerate lowest unoccupied molecular 
orbital (LUMO) and is delocalized over the whole molecule. The analysis
of the particle-hole pairs needed to describe these excitations  
show an important contribution coming from the  particle-hole pair with
negative energy
at minus the HOMO-LUMO gap, that is neglected within the TDA. 
These results confirm the trend of blue-shifting and overestimating
the intensity of excited state with a more delocalized character (such
as $\pi\rightarrow \pi^*$ excitations in molecules)
 that has been already observed and discussed in the literature.~\cite{GruningMG09,MaRM09}

\subsection{Accuracy of the pseudo-Hermitian Lanczos algorithm}~\label{sc:resacc} 
\begin{figure}[h]
\begin{center}
\includegraphics[clip,width=8cm]{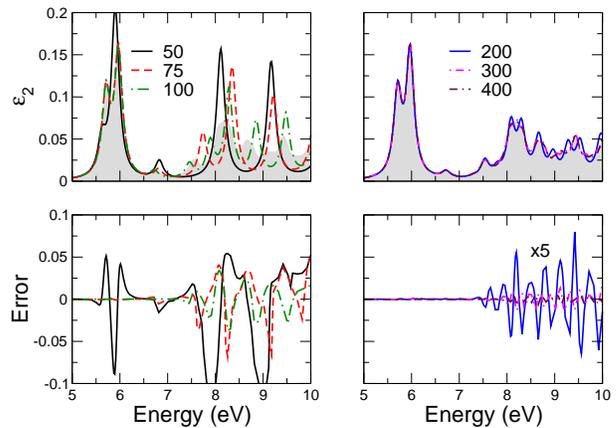}
\end{center}
\vspace{-.5cm}
\caption{[Color online] Top panels: Absorption spectrum of 1,3,5-TCB obtained by
  exact diagonalization (gray shadow) or by the LH iterative procedure
for a different number of iterations. Bottom panels: error in spectrum
obtained by the LH method with respect to exact diagonalization. Note
that in the right bottom panel the curve have been magnified by a
factor 5. 
The matrix elements in Eqs.~(\ref{eq:bsemte}) of the Hamiltonian have been
calculated with $W_{\alpha\beta,\lambda\mu}=0$  in Eq.~(\ref{eq:bsekrn}) and without quasiparticle corrections.}
\label{fg:AccPLH}
\end{figure}

At each iteration the LH algorithm provides an approximation for the
eigenvalue spectrum of the RPA Hamiltonian and thus, through the
macroscopic dielectric function, of the optical absorption spectrum.
Figure~\ref{fg:AccPLH} shows how by increasing the number of iterations, the spectrum
obtained by the LH approach becomes more and more accurate, until the
results are, for about 300 iterations, indistinguishable on the scale of the plot.
Note that the spectrum does not converge uniformly for all energies,
but it converges before in the low energy part, and then progressively
for higher excitations. In fact if we restrict ourselves to the part of the spectrum 
examined in the experiment would need only about 75 iterations. 
%%%
%%%

\begin{figure}[h]
\begin{center}
\includegraphics[clip,width=8cm]{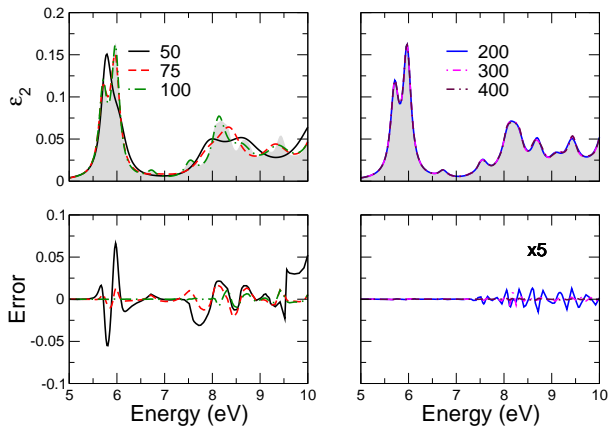}
\end{center}
\vspace{-.5cm}
\caption{As Fig.~\ref{fg:AccPLH}, but the terminator in Eq.~(\ref{eq:termgp}) has been used for
  the continued fraction in Eqs.~(\ref{eq:cntfrc},~\ref{eq:cntfr2}).}
\label{fg:APLHTr}
\end{figure}

The accuracy can be
improved by properly terminating the continued fraction in
Eqs.~(\ref{eq:cntfrc},~\ref{eq:cntfr2}). For the spectra in
Fig.~\ref{fg:AccPLH} we just truncated the continued fraction.   
Instead, as suggested in Ref.~\onlinecite{Rocca08}, in
Fig.~\ref{fg:APLHTr} the asymptotic behavior of the continued
fraction is described by the terminator
\begin{equation}\label{eq:termgp}
g(\omega) = \frac{\omega^2 + b_{\text{u}}^2 - b_{\text{g}}^2 +\sqrt{(\omega^2 +
    b_\text{u}^2 - b_\text{g}^2)^2 - 4\omega^2 b_{\text{u}}^2}}{2\omega b_{\text{u}}^2},
\end{equation}
that is obtained by resumming the continued fraction corresponding to a
tight-binding Hamiltonian with the hopping parameters 
oscillating between two values, $b_{\text{u}}$ and
$b_{\text{g}}$.~\cite{endnote31}
For $b_{\text{u}}$ and $b_{\text{g}}$ we use the averages of the
odd and even $b_i$ in Eq.~(\ref{eq:tridia}) respectively.
In fact we have verified that in Eq.~(\ref{eq:tridia}) the odd (even) $b_i$
oscillate around its asymptotic value $b_{\text{u}}$
($b_{\text{g}}$).~\cite{endnote32}

The terminator improves the quality of the spectrum obtained by
iteration especially for higher energies (as it should be since we are
correcting the asymptotic behavior) where it partially eliminates the
spurious oscillations due to the truncation of the continued
fraction. Using the terminator we then reduce to 200 the number of
iterations needed for satisfactorly reproducing the exact
diagonalization results 
in the whole energy range we considered. 
%%%
%%% 
\subsection{Efficiency of the pseudo-Hermitian Lanczos algorithm}~\label{sc:reseff}  
\begin{figure}[h]
\begin{center}
\epsfig{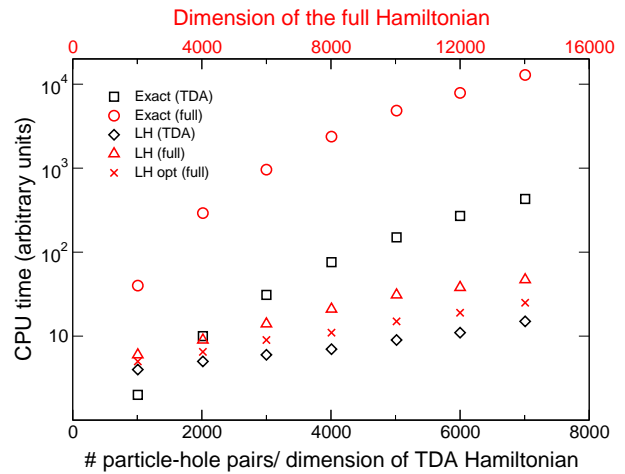}
\end{center}
\vspace{-.5cm}
\caption{Time needed to diagonalize the Hamiltonian
  against the number of particle-hole pairs, 
  either by exact diagonalization (using the
  {\sc lapack} from the Intel Mathematical Kernel Library) or by the
LH method (200 iterations, 1500 frequencies in the range between 0-15 eV). For the
TDA Hamiltonian, the LH in Fig.~(\ref{fg:lnczsH}) has been
used. For the full Hamiltonian both the pseudo-Hermitian LH
in Fig.~(\ref{fg:lncPTH}) and the optimized pseudo-Hermitian LH in Fig.~(\ref{fg:lncEMH}) have been used. 
The CPU time (one core on an Intel/Xeon) has been measured using standard C
timing routines.}
\label{fg:timPLH}
\end{figure}

The computational time is determined by two
factors: the dimension of the Hamiltonian and the number of iterations
needed to reach a desired accuracy. 
First, we analyze the behavior of the algorithm with respect to the dimension of the Hamiltonian. 
Figure~\ref{fg:timPLH} compares the timing for exact and iterative diagonalization of the BSE
using the LH
approach, either including (pseudo-Hermitian $H$) or neglecting
(Hermitian $H$) the coupling. The number of particle-hole pairs has been gradually increased so to
enlarge the dimension of the Hamiltonian. Note that for the TDA
Hamiltonian the dimension is equal to the particle-hole pairs, while
for the full Hamiltonian  the dimension is equal to twice the particle-hole pairs.  
For the LH approach the number of iteration is kept fixed.

When the number of particle-hole pairs exceeds 1000, the iterative methods become faster than exact
 diagonalization: for about 7000 particle-hole pairs the LH iterative
 procedure is about two order of
 magnitude faster for the Hermitian case, and three order of magnitude
 faster for the pseudo-Hermitian case.
In fact, the computational time of diagonalization is increasing rapidly with
 the dimension of the matrix, it grows about two order of magnitude while
 the Hamiltonian grows by about one order of magnitude.   
On the other hand the computational time of the LH approach is increasing slowly with
 the dimension of the matrix, while the Hamiltonian dimension grows of
 one order of magnitude, the computation time remains of the same order of
 magnitude.
Note that the non-Hermicity has large impact on the timing of the
 diagonalization, while--thanks to the algorithm presented in Fig.~\ref{fg:lncEMH}--it does not influence
 the performance of the iterative method considering that the full
 Hamiltonian has twice the dimensions if the TDA one.
%%%
%%%

\begin{figure}[h]
\begin{center}
\includegraphics[clip,width=8cm]{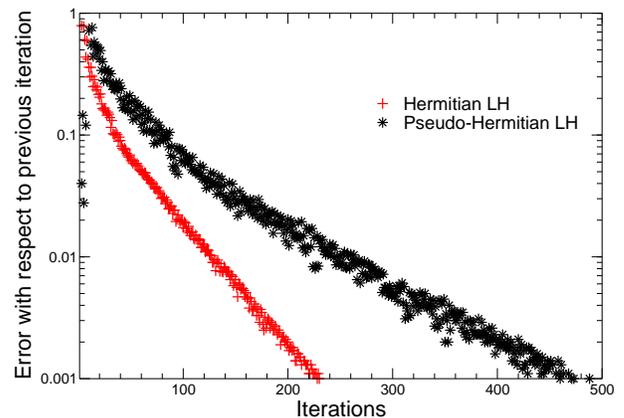}
\end{center}
\vspace{-.5cm}
\caption{
Fixed the Hamiltonian dimension (N=4824), 
convergence behavior with the number of iterations of the Hermitian and pseudo-Hermitian LH.}
\label{fg:EffPLH}
\end{figure}

Second, we turn to the number of iterations
needed to fulfill a given convergence criterium. 
Figure~\ref{fg:EffPLH} compares, fixed the dimension of the
Hamiltonian, the error on the spectrum with the
number of iteration in the Hermitian and pseudo-Hermitian case. We measured 
the error at each iteration $j$ 
as the frequency-weighted sum (on all the frequencies within the considered energy range)
of the differences with respect to the previous iteration $\sum_i
|S_i(j)-S_i(j-1)|/S_i(j)/\omega_i$, where $S_i$ is the value of the
spectrum at $\omega_i$.~\cite{endnote33}  
%%%
%%%
It appears that the LH approach applied to the Hermitian TDA
Hamiltonian converges to a given threshold about twice as fast than the
LH approach for the full Hamiltonian.
In order to explore the causes of the slower convergence for the full
Hamiltonian, we have considered the Hermitian Hamiltonian 
\begin{equation}
H^{\text{no coupl}} =\begin{pmatrix} R & 0\\ 0 & -R^*\end{pmatrix}
\end{equation}
obtained by setting the coupling
elements to zero in Eq.~(\ref{eq:rparpa}), and we have diagonalized it using both the
Hermitian and the pseudo-Hermitian LH algorithms. In both
cases we found the convergence rate is the same as for the full
Hamiltonian with the coupling. This indicates that neither the
coupling/non-Hermicity, neither the pseudo-Hermitian algorithm are responsible for the
slower convergence rate. 
We conjecture that the difference is due to different spectrum of the
TDA  Hamiltonian and the $H^{\text{no coupl}}$.
In Ref.~\onlinecite{ChenGuo}, Chen and Guo have shown for the simple Lanczos
that fixed the number of iterations $k$,
the number of converged eigenvalues $m$ in the Lanczos method was inversely proportional to
the spectral range $r$ of the Hamiltonian (defined as $r=E_{\text{max}}-E_{\text{min}}$, where
$E_{\text{max/min}}$ is the largest/smallest eigenvalue of H).  
Considering our particular case, the energy spectrum of $R$ in
$H^{\text{no coupl}}$ goes from (approximative estimate from
independent particle energies) 4.4 eV to 27 eV, instead the energy
spectrum of $R'$ (of the same dimension $N$ as $H^{\text{no coupl}}$)
goes from 4.4 eV to 28 eV~\cite{endnote34}
So practically the energy range of $R$ and $R'$ is very similar, and
the energy range of $H^{\text{no coupl}}$ goes from -27 eV to 27 eV,
that is about twice as large than that of $R'$. This may explain the slower
convergence rate of the LH method for $H^{\text{no coupl}}$.

\section{Summary}
In this work we have explained the theoretical ground and detailed the
derivation of the pseudo-Hermitian Lanczos recursion method for RPA
Hamiltonian introduced in
Ref.~\onlinecite{GruningMG09}. We also discussed, using numerical examples,
the accuracy and the computational cost of the method when compared
with conventional diagonalization techniques and the Hermitian Lanczos
recursion method. As expected the cost per iteration of the Hermitian
and pseudo-Hermitian (at parity of matrix dimension) is the same. On
the other hand the number of iterations needed to reach a desired
accuracy in the Lanczos eigenvalues spectrum is larger (about the
double) of the Hermitian case. We have tested that this behavior does
not depend on either the non-Hermicity of the Hamiltonian or the
pseudo-Hermitian algorithm: we argue that it is caused by the larger
spectral range. In fact the RPA spectra includes both negative and
positive particle-hole pairs, thus it is about the double of the TDA
spectra. 
We expect the pseudo-Hermitian LH algorithm for the RPA to have
particular relevance within computational condensed matter physics and
theoretical chemistry, where the calculation of the optical response of
materials require the solution of large eigenproblems for RPA
Hamiltonian. In fact, as shown briefly in this work, and
discussed in Ref.~\onlinecite{OlevanoR01,GruningMG09,MaRM09}, the
treatment of the full RPA Hamiltonian is important for accurately
reproducing optical spectra of molecules, nanostructures and the
energy loss function in solids.
Moreover by using the pseudo-Hermicity property it is possible in principle to
conveniently reformulate any Hermitian algorithm for RPA
eigenproblems.

\section*{Acknowledgments}
M.~G. acknowledges the usage of computer resources at the Laboratory
for Advanced Computing of the University of Coimbra and
is grateful to the Funda\c{c}\~{a}o para a Ci\^{e}ncia e a Tecnologia (FCT) for its support through the Ci\^{e}ncia 2008 programme.
A.~M. thanks the HPC-Europa2 Transnational mobility program 
(RII3-CT-2003-506079).
X.~G. acknowledges support from the Belgian State - Belgian Science Policy through the Interuniversity Attraction Poles Program (P6/42), the 
EU's 7th Framework Programme through the ETSF I3 e-Infrastructure project (Grant Agreement 211956), 
the Walloon region Belgium (RW project N${}^{\circ}$ 816849, WALL-ETSF), FRS-FNRS Belgium (FRFC Grant N${}^{\circ}$ 2.4.589.09.F), 
the Agentschap voor Innovatie door Wetenschap en Technologie (IWT project N${}^{\circ}$ 080023, ISIMADE),
and the Communaut\'e fran\c{c}aise de Belgique through the NANHYMO project (ARC 07/12-003).

\appendix
\section{Off-diagonal resolvent matrix elements in the Lanczos
  basis}\label{ap:offdia}
The off-diagonal resolvent matrix elements in the Lanczos basis at
iteration $j$,  
\begin{equation}\label{eq:offdia}
G_{i,0}^j = \langle q_i|(\omega -H)^{-1} |u_0\rangle,
\end{equation}  
can be obtained by recursion considering the system of linear equation 
\begin{equation}\label{eq:greens}
(\omega I - T^j) G^j_{i,0} = \delta_{i0},
\end{equation}
with $T^j$ defined in Eq.~(\ref{eq:tridia}) from which it follows
\begin{eqnarray} 
G^j_{1,0} &=& (1 - (\omega - a_1) G^j_{0,0})/b_2, \\
G^j_{n,0} &=& (-b_{n}G^j_{n-2,0} - (\omega - a_{n}) G^j_{n-1,0})/b_{n+1}.\label{eq:offdi1}
\end{eqnarray}
Alternatively, by direct solution of Eq.~(\ref{eq:greens}) one
obtains
\begin{equation}
G^j_{n,0} = (-1)^n \Pi_{i=1}^{n} b_{i+1} \frac{|A^j_{n+1}|}{|A^j_{0}|}.
\end{equation} 
where $A^j= \omega I - T^j $, and $A^j_i$ is the minor of $A^j$
obtained by deleting the first $i$ rows and columns.
This expression can be rewritten to obtain a recursion formula as,
\begin{equation}\label{eq:offdi2}
G^j_{n,0} = -b_{n+1} \frac{|A^j_{n+1}|}{|A^j_{n}|} G^j_{n-1,0}.
\end{equation} 
Equations~(\ref{eq:offdi2}) and~(\ref{eq:offdi1}) are
equivalent [using $G^j_{00}
  = |A^j_1|/|A^j_0|$ and the relation $|A^j_i| = (\omega - a_{i+1})
  |A^j_{i+1}| - b^2_{i+2}|A^j_{i+2}|$ in Eq.~(\ref{eq:offdi1}), one
  obtains Eq.~(\ref{eq:offdi2})], but they may give different results
due to the propagation of numerical error. In fact, since
Eq.~(\ref{eq:offdi1}) depends on $\omega$, the 
error $\delta$ on $G^j_{0,0}$ propagates and affects  $G^j_{n,0}$ by about $(\omega)^{n+1}
\delta$. Eq.~(\ref{eq:offdi2}) instead, the error on $G^j_{n,0}$ is proportional to
$(|A_{n+1}|/|A_{1}|)\delta$. From numerical test we have seen that
Eq.~(\ref{eq:offdi1}) gives indeed numerical problems for large $n$,
thus Eq.~(\ref{eq:offdi2}) has been used to calculate the
off-diagonal resolvent matrix elements in Eq.~(\ref{eq:cntfrP}).
Since $|A^j_{n+1}|/|A^j_{n}|$ are computed
already to obtain  $G^j_{0,0}$, Eq.~(\ref{eq:offdi2}) does not introduce extra
computational cost.

%%%
%%%%%%%%%%%%%%%%%%%%%%%%%%%%%%%%%%%%%%%%%
%%%%%%%%%%%%%%%%%%%%%%%%%%%%%%%%%%%%%%%%%
\bibliographystyle{apsrev}

\begin{thebibliography}{22}
\expandafter\ifx\csname natexlab\endcsname\relax\def\natexlab#1{#1}\fi
\expandafter\ifx\csname bibnamefont\endcsname\relax
  \def\bibnamefont#1{#1}\fi
\expandafter\ifx\csname bibfnamefont\endcsname\relax
  \def\bibfnamefont#1{#1}\fi
\expandafter\ifx\csname citenamefont\endcsname\relax
  \def\citenamefont#1{#1}\fi
\expandafter\ifx\csname url\endcsname\relax
  \def\url#1{\texttt{#1}}\fi
\expandafter\ifx\csname urlprefix\endcsname\relax\def\urlprefix{URL }\fi
\providecommand{\bibinfo}[2]{#2}
\providecommand{\eprint}[2][]{\url{#2}}

\bibitem[{\citenamefont{Fetter and Walecka}(1971)}]{FetterW71}
\bibinfo{author}{\bibfnamefont{A.~L.} \bibnamefont{Fetter}} \bibnamefont{and}
  \bibinfo{author}{\bibfnamefont{J.~D.} \bibnamefont{Walecka}},
  \emph{\bibinfo{title}{Quantum Theory of many-particle systems}}
  (\bibinfo{publisher}{Dover}, \bibinfo{year}{2003}),
  chap.~\bibinfo{chapter}{15}, p. \bibinfo{pages}{565}.

\bibitem[{\citenamefont{Ring and Schuck}(1980)}]{RingS80}
\bibinfo{author}{\bibfnamefont{P.} \bibnamefont{Ring}} \bibnamefont{and}
  \bibinfo{author}{\bibfnamefont{P.} \bibnamefont{Schuck}},
  \emph{\bibinfo{title}{The Nuclear Many-Body Problem (Theoretical and Mathematical Physics)}}
  (\bibinfo{publisher}{Springer}, \bibinfo{year}{1980}),
  chap.~\bibinfo{chapter}{7,8}, p. \bibinfo{pages}{244}.

\bibitem[{\citenamefont{Runge and Gross}(1984)}]{RungeG84}
\bibinfo{author}{\bibfnamefont{E.}~\bibnamefont{Runge}} \bibnamefont{and}
  \bibinfo{author}{\bibfnamefont{E.~K.~U.} \bibnamefont{Gross}},
  \bibinfo{journal}{Phys. Rev. Lett.} \textbf{\bibinfo{volume}{52}},
  \bibinfo{pages}{997} (\bibinfo{year}{1984}).

\bibitem[{\citenamefont{Salpeter and Bethe}(1951)}]{SalpeterB51}
\bibinfo{author}{\bibfnamefont{E.~E.} \bibnamefont{Salpeter}} \bibnamefont{and}
  \bibinfo{author}{\bibfnamefont{H.~A.} \bibnamefont{Bethe}},
  \bibinfo{journal}{Phys. Rev.} \textbf{\bibinfo{volume}{84}},
  \bibinfo{pages}{1232} (\bibinfo{year}{1951}).

\bibitem{endnote27}{Because of the 4-point nature of the kernel, within BS the reformulation in terms of the eigenproblem for $H^{\protect \text  {RPA}}$ is the only feasible alternative. To the contrary, within TD-DFT the direct solution of the Dyson-like equation is affordable and even the most convenient in the case of bulk periodic systems. On the other hand, for finite or sparse systems even within TDDFT the solution of the eigenproblem for ${\protect \bf  H}^{\protect \text  {RPA}}$ can be more efficient than the direct solution especially when one is interested only to a frequency range where the excitations are well separated.}

\bibitem[{\citenamefont{Onida et~al.}(2002)\citenamefont{Onida, Reining, and
  Rubio}}]{OnidaRR02}
\bibinfo{author}{\bibfnamefont{G.}~\bibnamefont{Onida}},
\bibnamefont{et al.},
  \bibinfo{journal}{Rev. Mod. Phys.} \textbf{\bibinfo{volume}{74}},
  \bibinfo{pages}{601} (\bibinfo{year}{2002}).

\bibitem[{\citenamefont{Bai et al}(2000)}]{Bai00}
\emph{\bibinfo{booktitle}{Templates for the solution of algebraic
    problems: a practical guide}}, edited by
\bibinfo{editor}{\bibfnamefont{Z.} \bibnamefont{Bai}},
\bibinfo{editor}{\bibfnamefont{J.} \bibnamefont{Demmel}},
\bibinfo{editor}{\bibfnamefont{J.} \bibnamefont{Dongarra}},
\bibinfo{editor}{\bibfnamefont{A.} \bibnamefont{Ruhe}},
\bibinfo{editor}{\bibfnamefont{H.} \bibnamefont{van der Vorst}},
   (\bibinfo{publisher}{SIAM}, \bibinfo{address}{Philadelphia},
   \bibinfo{year}{2000}).

\bibitem{Tretiak09}
S. Tretiak, C. M. Isborn, A. M. N. Niklasson, and M. Challacombe
J. Chem. Phys. {\bf 130}, 054111 (2009)

\bibitem[{\citenamefont{Casida}(1995)}]{Casida95}
\bibinfo{author}{\bibfnamefont{M.}~\bibnamefont{Casida}}, in
  \emph{\bibinfo{booktitle}{Recent Advances in Density Functional Methods}},
  edited by \bibinfo{editor}{\bibfnamefont{D.~P.} \bibnamefont{Chong}}
  (\bibinfo{publisher}{World Scientific}, \bibinfo{address}{Singapore},
  \bibinfo{year}{1995}), vol.~\bibinfo{volume}{1}.

\bibitem{OlevanoR01} V. Olevano and L. Reining, Phys. Rev. Lett. {\bf 86},
  5962 (2001).

\bibitem{MarinopolousG10} A. Marinopoulos and M. Gr\"uning, in preparation.

\bibitem{GruningMG09} M. Gr\"uning, A. Marini and X. Gonze, Nano Lett. {\bf 9},
  2820 (2009).

\bibitem{MaRM09} Y. Ma, M. Rohlfing and C. Molteni, Phys. Rev. B {\bf 80},
  241405 (2009).

\bibitem[{\citenamefont{Van der Vorst}(1982)\citenamefont{H. A. van der Vorst}}]{VdVorst82}
H. A. van der Vorst Math. Comp. 39 (1982), 559 

\bibitem{Grosso95} Note that in the Lanczos approach the energy range
  includes always the lowest energies. In principle it is possible to
  choose a different energy range, not necessarily containing the
  lowest energy, using the modification proposed by  
G. Grosso, L. Martinelli and G. Pastori Parravini, Phys. Rev. B {\bf 51}, 13033–13038 (1995) 

\bibitem{Haydock80}
 R. Haydock, in
 \emph{Solid State Phys.}, {\bf 35} 215 (1980)
 edited by
 H. Ehrenfest, F. Seitz, and D. Turnbull, Academic Press. 

\bibitem{Cini07}
 M. Cini, \emph{Topics and Methods in Condensed Matter Theory
From Basic Quantum Mechanics to the Frontiers of Research} (Springer,
 Berlin, 2007) p. 313-316.

\bibitem{endnote28}{In particular, at each iteration both the computational cost and the storage are increasing, since all the vectors in the basis are needed to determine the new one.}

\bibitem{endnote29}{In principle also the Hermitian algorithm can break down. In fact after many iterations, because of the numerical error, the vectors in the Lanczos basis may stop to be orthonormal.}
\bibitem[{\citenamefont{Walker et~al.}(2006)\citenamefont{Walker, Saitta,
  Gebauer, and Baroni}}]{Walker06}
\bibinfo{author}{\bibfnamefont{B.}~\bibnamefont{Walker}},
  \bibinfo{author}{\bibfnamefont{A.~M.} \bibnamefont{Saitta}},
  \bibinfo{author}{\bibfnamefont{R.}~\bibnamefont{Gebauer}}, \bibnamefont{and}
  \bibinfo{author}{\bibfnamefont{S.}~\bibnamefont{Baroni}},
  \bibinfo{journal}{Phys. Rev. Lett.} \textbf{\bibinfo{volume}{96}},
  \bibinfo{pages}{113001} (\bibinfo{year}{2006}).

\bibitem{Rocca08}
D. Rocca, R. Gebauer, Y. Saad, and S. Baroni, J. Chem. Phys. {\bf 128},
154105 (2008).

\bibitem{Rocca10}
D. Rocca, D. Lu, and G. Galli, J. Chem. Phys. {\bf 133}, 164109 (2010).

\bibitem[{\citenamefont{Mostafazadeh}(2002{\natexlab{b}})}]{Mostafazadeh02a}
\bibinfo{author}{\bibfnamefont{A.}~\bibnamefont{Mostafazadeh}},
  \bibinfo{journal}{J. Math. Phys.} \textbf{\bibinfo{volume}{43}},
  \bibinfo{pages}{205} (\bibinfo{year}{2002}{\natexlab{b}}).

\bibitem[{\citenamefont{Mostafazadeh}(2002{\natexlab{a}})}]{Mostafazadeh02b}
\bibinfo{author}{\bibfnamefont{A.}~\bibnamefont{Mostafazadeh}},
  \bibinfo{journal}{J. Math. Phys.} \textbf{\bibinfo{volume}{43}},
  \bibinfo{pages}{3944} (\bibinfo{year}{2002}{\natexlab{a}}).

\bibitem[{\citenamefont{Zimmermann}(1970)}]{Zimmermann70}
\bibinfo{author}{\bibfnamefont{R.}~\bibnamefont{Zimmermann}},
  \bibinfo{journal}{Phys. Stat. Sol.} \textbf{\bibinfo{volume}{41}},
  \bibinfo{pages}{23} (\bibinfo{year}{1970}).

\bibitem{BenedictS99}
L. X. Benedict and E. L. Shirley, Phys. Rev. B {\bf 59}, 5441 (1999).

\bibitem{endnote30}{For real matrices, as in the case of TD-DFT for finite systems and local exchange-correlation functionals, the cost is reduced further, since one can construct before-hand the two matrix $A=R+C$ and $B=R-C$, so that only a single matrix-vector multiplication is needed (giving another factor two). Moreover usually $B$ reduces to a diagonal matrix, containing on the diagonal the energy differences of the particle-hole pairs. In this case the algorithm is equivalent to the one proposed by Casida, see Ref.~\onlinecite{Casida95}.}

\bibitem[{\citenamefont{Gonze \emph{et~al.}}(2002)\citenamefont{Gonze,
  Beuken, Caracas, Detraux, Fuchs, Rignanese, Sindic, Verstraete, Zerah,
  Jollet \emph{et~al.}}}]{abinitref}
\bibinfo{author}{\bibfnamefont{X.}~\bibnamefont{Gonze}},
  \bibinfo{author}{\bibfnamefont{J.~M.} \bibnamefont{Beuken}},
  \bibinfo{author}{\bibfnamefont{R.}~\bibnamefont{Caracas}},
  \bibinfo{author}{\bibfnamefont{F.}~\bibnamefont{Detraux}},
  \bibinfo{author}{\bibfnamefont{M.}~\bibnamefont{Fuchs}},
  \bibinfo{author}{\bibfnamefont{G.~M.} \bibnamefont{Rignanese}},
  \bibinfo{author}{\bibfnamefont{L.}~\bibnamefont{Sindic}},
  \bibinfo{author}{\bibfnamefont{M.}~\bibnamefont{Verstraete}},
  \bibinfo{author}{\bibfnamefont{G.}~\bibnamefont{Zerah}},
  \bibinfo{author}{\bibfnamefont{F.}~\bibnamefont{Jollet}},
  \bibnamefont{\emph{et~al.}}, \bibinfo{journal}{Comp. Mater. Sci.}
  \textbf{\bibinfo{volume}{25}}, \bibinfo{pages}{478} (\bibinfo{year}{2002});
\bibinfo{author}{\bibfnamefont{X.}~\bibnamefont{Gonze}},
  \bibinfo{author}{\bibfnamefont{G.~M.} \bibnamefont{Rignanese}},
  \bibinfo{author}{\bibfnamefont{M.}~\bibnamefont{Verstraete}},
  \bibinfo{author}{\bibfnamefont{J.~M.} \bibnamefont{Beuken}},
  \bibinfo{author}{\bibfnamefont{Y.}~\bibnamefont{Pouillon}},
  \bibinfo{author}{\bibfnamefont{R.}~\bibnamefont{Caracas}},
  \bibinfo{author}{\bibfnamefont{F.}~\bibnamefont{Jollet}},
  \bibinfo{author}{\bibfnamefont{M.}~\bibnamefont{Torrent}},
  \bibinfo{author}{\bibfnamefont{G.}~\bibnamefont{Zerah}},
  \bibinfo{author}{\bibfnamefont{M.}~\bibnamefont{Mikami}},
  \bibnamefont{\emph{et~al.}}, \bibinfo{journal}{Z. Kristallogr.}
\textbf{\bibinfo{volume}{220}}, \bibinfo{pages}{558} (\bibinfo{year}{2005})
  \bibinfo{author}{\bibfnamefont{X.}~\bibnamefont{Gonze}},
  \bibinfo{author}{\bibfnamefont{B.} \bibnamefont{Amadon}},
  \bibinfo{author}{\bibfnamefont{P.~M.}~\bibnamefont{Anglade}},
  \bibinfo{author}{\bibfnamefont{J.~M.} \bibnamefont{Beuken}},
  \bibinfo{author}{\bibfnamefont{F.}~\bibnamefont{Bottin}},
  \bibinfo{author}{\bibfnamefont{P.}~\bibnamefont{Boulanger}},
  \bibinfo{author}{\bibfnamefont{F.}~\bibnamefont{Bruneval}},
  \bibinfo{author}{\bibfnamefont{D.}~\bibnamefont{Caliste}},
  \bibinfo{author}{\bibfnamefont{R.}~\bibnamefont{Caracas}},
  \bibinfo{author}{\bibfnamefont{M.}~\bibnamefont{C\^ot\'e}},
  \bibnamefont{\emph{et~al.}}, \bibinfo{journal}{Comp. Phys. Commun.}
  \textbf{\bibinfo{volume}{180}}, \bibinfo{pages}{2582} (\bibinfo{year}{2009}).
\bibitem[{\citenamefont{Marini}(2006-2008)}]{yamboref}
\bibinfo{author}{\bibfnamefont{A.}~\bibnamefont{Marini}},
\bibinfo{author}{\bibfnamefont{C.}~\bibnamefont{Hogan}},
\bibinfo{author}{\bibfnamefont{M.}~\bibnamefont{Gr\"uning}}
\bibnamefont{and}
\bibinfo{author}{\bibfnamefont{D.}~\bibnamefont{Varsano}},
  \bibinfo{journal}{Comp. {P}hys. {C}omm.} \textbf{\bibinfo{volume}{180}},
  \bibinfo{pages}{1392} (\bibinfo{year}{2009}).

\bibitem{Sharping87}
Absorption measurements at a resolution of 50 cm$^{-1}$ using a gas
saturation method (298 K) by 
H. Scharping, C. Zetzsch, and H. A. Dessouki,  J. Mol. Spectrosc. {\bf 123}, 382 (1987).
\bibitem{ChenGuo}
R. Chen, and H. Guo, Chem. Phys. Lett {\bf 369}, 650 (1997);
ibid. J. Chem. Phys. {\bf 119}, 5762 (2003).  
\bibitem{endnote31}{This corresponds to a two-bands system, with bands between $|b_{\protect \text  {u}} - b_\protect \text  {g}|$ and $(b_{\protect \text  {u}} + b_\protect \text  {g})$, and between $-(b_{\protect \text  {u}} + b_\protect \text  {g})$ and $-|b_{\protect \text  {u}} - b_\protect \text  {g}|$.}
\bibitem{endnote32}{The sum of the asymptotic values tends to twice the particle-hole pair energy cutoff, the difference to the optical gap of the system. See also Ref.\protect \nobreakspace  {}\protect \onlinecite  {Rocca08}.}
\bibitem{endnote33}{We have tested that the convergence behavior does not depend on the particular expression for the error and that the behavior resemble that of the error with respect to the exact diagonalization, at least when no terminator is used.}
\bibitem{endnote34}{The reason of just 1 eV difference, while the unoccupied orbital space of $R'$ contains 25 more states, is that KS states in this energy region are in fact very close to each others.}


\end{thebibliography}

\end{document}